\documentclass[fleqn,usenatbib,letters]{mnras}

\usepackage{graphicx}	
\usepackage{amsmath}	
\usepackage{amssymb}	

\title[Radio emission from PTF10hgi]{Radio emission from a decade old Type I superluminous supernova, PTF10hgi: Comparison with FRB121102}

\author[Surajit Mondal, Apurba Bera, Poonam Chandra, Barnali Das]{Surajit Mondal$^{1}$\thanks{Contact e-mail: \href{mailto:surajit@ncra.tifr.res.in}{surajit@ncra.tifr.res.in}},Apurba Bera$^{1}$, Poonam Chandra$^{1}$, Barnali Das$^{1}$
\\
$^{1}$National Centre for Radio Astrophysics, Tata Institute of Fundamental Research, Pune, India}

\date{\today}

\pubyear{2020}

\begin{document}
\label{firstpage}
\pagerange{\pageref{firstpage}--\pageref{lastpage}}
\maketitle













\begin{abstract}
{
We perform a comparative study between the only radio detected Type I superluminous supernova (SLSN) PTF10hgi, and the most active repeating fast radio burst FRB121102. This study has its root in the hypothesized FRB-SLSN connection that states that magnetars born in SLSN can power FRBs. The wideband spectrum (0.6--15 GHz) of PTF10hgi presented here, provides strong evidence for the magnetar wind nebular origin of the radio emission. The same spectrum also enables us to make robust estimates of the radius and the magnetic field of the radio emitting region and demonstrates that the nebula is powered by the rotational energy of the magnetar. This spectrum is then compared with that of FRB121102 which we extend down to 400 MHz using archival data. The newly added measurements put very tight constraint on the emission models of the compact persistent source associated with FRB121102. We find that while both sources can be powered by the rotational energy of the underlying magnetar, the average energy injection rate is much higher in FRB121102. Hence, we hypothesise that, if PTF10hgi is indeed emitting fast radio bursts, those will be much weaker energetically than those from FRB121102.

}
\end{abstract}



\section{Introduction}

Superluminous supernovae (SLSNe) are a type of supernovae which have an optical absolute magnitude $<-21$ and are more than 10 times brighter than typical supernovae. Similar to the typical supernovae, these are divided onto two broad classes based on the presence of hydrogen in their peak optical spectra. The SLSNe which lack the presence of strong hydrogen lines in their peak optical spectra are called SLSNe-I, while others are called SLSNe-II. 

SLSNe-II show multiple signs of interaction of circumstellar medium (CSM) and are assumed to be powered by the continued conversion of the kinetic energy of the supernova ejecta into the radiation by CSM interaction  \citep[e.g.][]{smith2007, chevalier2011}. Although such a scenario has also been proposed as the progenitor of SLSNe-I \citep[e.g.][]{sorokina2016}, there are competing models which suggest that at least some SLSNe-I are powered by a central engine, either a black hole accretion disc system \citep{dexter2013} or fast spinning newborn pulsar \citep[e.g.][]{kasen2010, woosley2010}, while other works suggest that these can also be powered by the radioactive decay of $^{56}$Co \citep{gal2009, young2010}. There have also been suggestions that both the lightcurve and luminosity of SLSNe-I can be explained by the so-called hybrid models which combine two or more of the above scenarios \citep[e.g.][]{moriya2015,tolstov2017}. 

Radio observations have the potential to distinguish between the different scenarios.
The radio emission being synchrotron in natures requires the presence of relativistic electrons and magnetic field. Hence detection of radio emission from a SLSNe will strongly suggest that the SLSNe luminosity had some contribution either from a pulsar/magnetar or due to interaction from the dense CSM. In the interaction picture, a steep spectrum is expected due to the synchrotron emission from the shock, whereas, a flat and optically thin spectrum can be interpreted to be originating from a pulsar/magnetar wind nebula \citep{reynolds2017}. 
{However, based on the properties of the confirmed pulsar wind nebulae (PWN) discovered till now, a very small fraction of PWNs can also have steep spectrum \citep{reynolds2017}. Hence, while a steep spectrum can both be, in principle, produced in the magnetar/synchrotron origin, a flat spectrum is only produced in the magnetar scenario.}
In the very rare event like in case of SN 1986J \citep{bietenholz2017}, it might also be possible to detect radio emission both from the wind nebula and the shock, which gives further support to the hybrid models. 

In addition to providing information about the progenitor, detection of radio emission is also significant for an entirely different, yet relevant, aspect. Following the discovery of a radio source coincident with the repeating FRB FRB121102 \citep{chatterjee2017}, { it was proposed that FRB121102 is probably powered by a decades old magnetar born in a Type I SLSN and it was hypothesised that most FRBs might have similar origin. However, recently multiple FRBs have been discovered whose properties are very different from that of FRB121102 and their host galaxies are also very different from the ones typically hosting Type I SLSNe \citep[e.g.][etc.]{bannister2019,ravi2019,bhandari2020}. Currently, it is believed that at least a subset of the FRB population is powered by decades old magnetars born in SLSNe. The galactic FRB detected from the magnetar SGR 1935+2154 \citep{anderson2020} also strengthens the connection between FRBs and magnetars. Under this scenario, the recent detection of radio emission from a Type I SLSN, PTF10hgi \citep{eftekhari19, law2019}, is a supporting evidence for the FRB-SLSN connection. However a quantitative comparison between the radio emission and related properties of these two classes of sources is still lacking.}

Here we present the near-simultaneous, wideband radio spectrum (0.6--15 GHz) of the SLSN PTF10hgi, and then compare it with the spectrum of FRB121102. For the latter, we extend the radio spectrum down to 0.4 GHz by analyzing archival data for this object acquired with the upgraded Giant Metrewave Radio Telescope (uGMRT) in band 3 (300--500 MHz) and band 4 (550--900 MHz). This comparative study not only provides important insights about the source of radio emission from PTF10hgi, but also enable us to make prediction about the energetics of any FRB(s) that may be produced from PTF10hgi.

\section{Observations \& results} \label{sec:results}

\begin{figure*}
 \centering
     \includegraphics[trim={0.47cm 0.3cm 0.3cm 0.3cm},clip,scale=0.55]{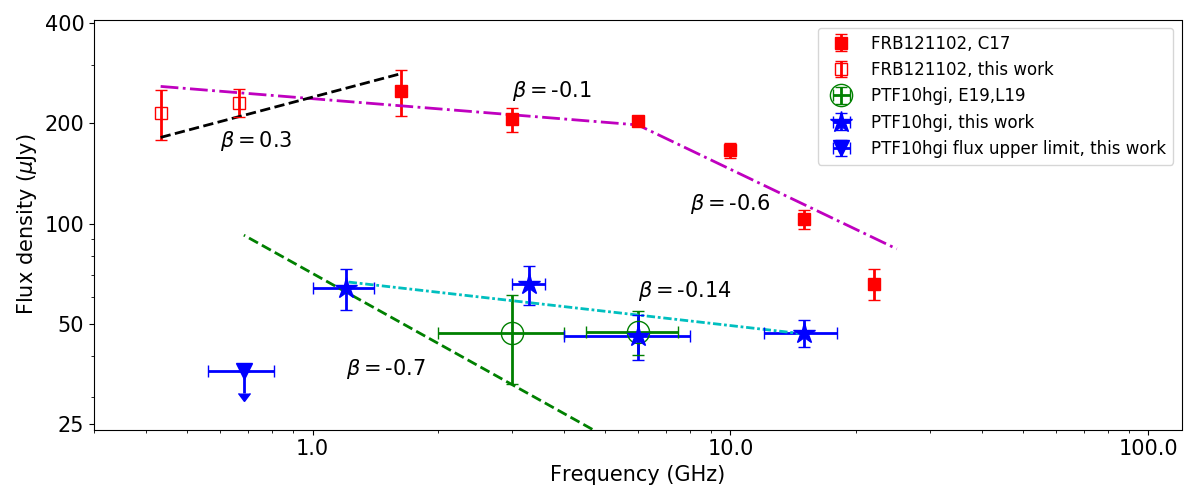}
     \caption{Wideband spectrum of PTF10hgi and compact source associated with FRB121102. 
     The triangle show the $3\sigma$ upper limit. 
     The magenta and cyan dashed-dotted lines are (broken) powerlaw fits. Powerlaw index ($\beta$) of each line is written near it. The black and green dashed lines are for indicative purposes only. E17: \citet{eftekhari19}, L19: \citet{law2019}, C17:\citet{chatterjee2017}}
     \label{fig:spectrum}
 \end{figure*}

{PTF10hgi was observed using the upgraded Giant Metrewave Radio Telescope (uGMRT) and Karl G. Jansky Very Large Array (VLA) at frequencies covering 0.6--15 GHz. The data for FRB121102 at frequencies 0.3--0.8 GHz were available from the uGMRT data archive. The details of the observations and data analysis are given in the Appendix. We also provide the images by zooming into the location of PTF10hgi in the Appendix. Here, we note that radio emission was detected from the location of PTF10hgi at all frequencies except at 0.6 GHz. The source is unresolved at all observed frequencies.} The spectra of the radio source associated with PTF10hgi and FRB121102 are shown in Fig. \ref{fig:spectrum}.
Below we note the following points from Fig. \ref{fig:spectrum}.

\begin{enumerate}
    \item 
    In spite of about a 2 year gap between the observations presented here and in \citet{eftekhari19} and \citet{law2019}, there is no significant difference in the flux density measurements for PTF10hgi at 3 and 6 GHz between the two epochs.
    
    \item A low-frequency break is clearly detected in case of PTF10hgi.
    Based on the steep drop in flux density at {0.6 GHz}, we infer that the synchrotron self-absorption frequency ($\nu_a$) is $\sim 1$ GHz.
    
    \item Between 1--18 GHz, the spectrum for PTF10hgi is very flat. The spectral index at these frequencies is $-0.15 \pm 0.06$.
    
    \item The spectrum of the persistent compact source associated with FRB121102 does not show any significant break between 0.4--6 GHz with 1.2 $\sigma$. The spectral index for FRB121102 between 0.4-6 GHz is $-0.1\pm 0.02$. This is very similar to the spectral index estimated for PTF10hgi between 1--18 GHz. However, the data are also consistent with increasing flux densities between {0.4--1 GHz}, although the spectral index is very shallow. The flux density measurements between {0.4--1 GHz} are consistent with a spectral index of $1/3$ as has been shown with the black dashed line. This might be indicative of the minimum injection frequency, $\nu_m\sim 1$ GHz and $\nu_a\lesssim$0.4 GHz.
    
    \item { The uGMRT band 3 observations for FRB121102 were taken in July--August 2017 about a year after the observations of \citet{chatterjee2017}. The lowest estimated age of the FRB121102 magnetar is $\gtrsim 10$ years \citep{margalit18a}. The radio emission from a decade old magnetar is not expected to change significantly in a year timescale (based on the radio emission from typical supernovae).  The uGMRT band 4 observations were taken on September 2019. Based on the fact that the flux density at 0.3 GHz and 0.6 GHz and other frequencies can be modelled well by a double powerlaw model, we conclude that there is no detectable flux density evolution in FRB121102 persistent source with time in our dataset.}

\end{enumerate}
\section{Discussion} \label{sec:discussion}

\subsection{Origin of the radio emission}\label{sec:origin_radio}


\citet{eftekhari19} suggested that radio emission of PTF10hgi is most probably originating from a magnetar wind nebula. However there were other possibilities like emission from interaction between the supernova ejecta and the circumstellar medium \citep{eftekhari19}, orphan afterglow of a long gamma ray burst (GRB) \citep{marcote2019} or an active galactic nuclei (AGN) origin, which could not be ruled out with the previously available data.

The wideband measurements reported here span a frequency range between 0.6--15 GHz. The source was detected at all frequencies at high SNR ($\gtrsim 5 \sigma$) expect at 0.6 GHz. Based on this, we derive that the spectral index of the emission between 1--15 GHz is $-0.15 \pm 0.06$, indicating that the source has a flat spectrum in the optically thin regime. The spectrum is evidently inconsistent with a spectral index of $-0.7$, which is typically expected for synchrotron emission from supernova ejecta and off-axis GRB. It is not clear how a flat spectrum might be produced in the optically thin regime in case of supernovae or off-axis jet after about a decade of explosion. Hence, these new data presented here effectively rules out these models. 

{
Flat spectrum radio sources are very often core dominated AGNs which are generally not found in dwarf galaxies.
Recently, \citet{reines2020} has discovered a number of compact radio sources in dwarf galaxies, which they hypothesise to be wandering black holes. The median spectral index of these sources is $-0.8$. There are only 2 sources out of their sample of 13 sources, which have a spectral index $<0.5$ in the optically thin regime. However, AGN nature of these sources cannot be confirmed with the current data. The properties of these sources are however consistent with the persistent source associated with FRB121102 \citep{eftekhari2020}, for which the magnetar model is the most probable model. 
}

The flat spectrum of the radio source associated with PTF10hgi can be easily produced by assuming that the radio emission is arising due to synchrotron emission from accelerated electrons which have a hard injection spectrum and are being accelerated by a magnetar/pulsar like central engine. \citet{law2019} predicted based on a fiducial model of the electron injection spectrum that there will not be any significant flux density variation at 3 and 6 GHz in a 2 year timescale, between 2017 and early 2020. Our measurements are also consistent with these predictions. Hence, the wideband flux density measurements of PTF10hgi reported here, provide the strongest evidence till date that the radio emission is originating from a pulsar/magnetar wind nebula, similar to that hypothesised to be behind the persistent radio emission from FRB121102 \citep{metzger2017,margalit2018a, margalit2018b}.  

\subsection{Physical parameters of PTF10hgi} \label{sec:PTF10hgi_properties}


We assume that the relativistic electrons responsible for the synchrotron emission can be modelled by a broken powerlaw model of the form \citep{law2019}. 

\begin{equation}
    N(\gamma)=
    \begin{cases}
    N_1\gamma^{-p_1}, \quad \gamma_{\rm{min}}<\gamma<\gamma_b,\\
    N_2 \gamma^{-p_2} \quad \gamma>\gamma_b,
    
    \end{cases}
    \label{eq:elec_dis}
\end{equation}

Here $\gamma$ is the Lorentz factor of the nonthermal electrons and $N(\gamma)$ is the density of nonthermal electrons with a Lorentz factor of $\gamma$. $\gamma_b$ is the break Lorentz factor and is dependent on the efficiency and energy content of the accelerator. $\gamma_{\rm{min}}$ is the minimum Lorentz factor of the accelerated electrons. $N_1$ and $N_2$ are the normalisation constants and are related to each other because of continuity at $\gamma=\gamma_b$. $p_1$ and $p_2$ are the particle energy indices. $p_1$ can be determined from the powerlaw fit to the spectrum between 1--15 GHz, since the spectral index is given by $(p_1-1)/2$. We have fitted a powerlaw model to the spectrum of PTF10hgi from 1--15 GHz and obtain $p_1=1.3$. We assume the same values of $\gamma_b=10^5$ and $p_2=2.5$ as taken by \citet{law2019}. For $p_1<2$, the exact value of $\gamma_{min}$ is not important for our calculations as $\gamma_b>>\gamma_{min}$. 

We assume that the self-absorption frequency, $\nu_a\sim 1\,$GHz, since the spectral index between 0.6--1.4 GHz is steeper than 2.5. 
The flux density observed in the optically thick part of the spectrum ($\nu<\nu_a$) is given by

\begin{equation}
    F (\nu)=\frac{\pi R^2}{D^2}\left( \frac{\nu}{\nu_a}\right)^{5/2}B^{-1/2} \times \left(\frac{c_{em}(p_1)}{c_{abs}(p_1)}+\frac{c_{em}(p_2)}{c_{abs}(p_2)}\right)
    \label{eq:nua_1}
    \end{equation}
    
   Flux density in the optically thin part ($\nu>\nu_a$) 
    \begin{equation}
    \begin{aligned}
    F(\nu)=&\frac{4\pi R^3f}{3D^2}\biggl(c_{em}(p_1)N_1B^{(p_1+1)/2}\left( \frac{\nu}{\nu_a}\right)^{(p_1-1)/2}+\\ & c_{em}(p_2)N_2B^{(p_2+1)/2}\left( \frac{\nu}{\nu_a}\right)^{(p_2-1)/2}       \biggr)
    \end{aligned}
    \label{eq:nua_2}
\end{equation}

Expressions $c_{em}$ and $c_{abs}$ are parameters dependent on $p$ and are provided in \citet{chevalier2017}. $R$ and $B$ are the radius and magnetic field of the radio emitting region. $D$ is the luminosity distance. $f$ is the { volume filling} factor. { The absolute values of $N_1$ and $N_2$ can be obtained in terms of the magnetic field and a parameter $\alpha$, where $\alpha$ is the ratio of the energy density in the accelerated electrons and the magnetic energy density.} Then we can solve eqs. \ref{eq:nua_1} and \ref{eq:nua_2} using $\nu_a\sim 1\,$GHz and $F(\nu_a)$ is equal to the flux density observed at 1.2 GHz. Under the conditions $f=1$ and $\alpha=1$, we obtain R=0.04 pc and B=0.02G. It has already been pointed out by \citet{chevalier1998} that these numbers are relatively robust to different values of $f$ and $\alpha$. Using the radius, we calculate that the wind velocity is $\sim 4000\,$km s$^{-1}$, which is similar to the typical speeds of wind nebula. { The magnetic field of 0.02 G estimated here is much higher than that typically found in the galactic pulsar wind nebulae, whose magnetic field is in the range $ 5\,\mu$G - $1\, m$G \citep{reynolds2012}. However, it is possible that there is an inherent difference between the magnetic field of pulsars/magnetars born from regular supernova and those born in SLSNe. The difference might also originate from the age difference between the galactic magnetars and PTF10hgi, which is only a decade old. }

The energy injected in the system is given by the magnetar is given by $ E_{\rm{tot}}=B^2R^3/(6\epsilon_B)$, where
$\epsilon_B$ is the fraction of energy in the magnetic field. 
Under the equipartition condition we calculate $E_{\rm{tot}}=2\times 10^{47}\,$erg. We have verified using different values of $\alpha$ and $f$ that the estimate of the energy content of the system is always of the order of $10^{47}\,$erg. This implies that the average energy injection rate of the magnetar is $3 \times 10^{38}\,$erg$\,$s$^{-1}$. To the best of our knowledge, this is the first robust determination of the energy content of the system and the average energy injection rate assuming the central engine scenario.

If the nebula is powered by the magnetic field of the magnetar, then the magnetic energy at birth should be at least $10^{47}\,$erg. However, using B=$3\times 10^{14}\,$G \citep{inserra13}, we calculate that the magnetic energy at birth should be $1.5\times 10^{46}\,$erg, assuming a radius of 10 km. This is about an order of magnitude lower than the fiducial energy content of the nebula. The uncertainties in the magnetic field estimated in \citet{inserra13} is small enough to conclude that this difference is real. Hence, we can safely conclude that the magnetar is not powering the nebula through its magnetic field. The second source of energy to be considered is the magnetar's rotational energy. We estimate its rotational energy considering it as a solid body with a mass of 1.4$M_\odot$, radius of 10 km and using a spin-period of $7.2$ms as given by \citet{inserra13}, and it comes out to be $2 \times 10^{50}$ erg, which is much larger than the total energy contained in the nebula. This implies that the nebula is rotation powered.

{Simulations suggest that the radio luminosity of the magnetar wind nebula decreases with time \citep{murase16}. The value of $\nu_a$ is also expected to decrease as the nebula expands and becomes optically thin to lower frequencies. Both of these predictions can be tested with future observations and will be very important in confirming this scenario. }


\subsection{Radio emission from FRB121102 and its physical properties}\label{sec:FRB121102}

Till date, there has been several works which try to explain the radio emission from FRB121102 in a quantitative manner. While all of these models agree that the magnetar wind nebula best explains the radio emissions, they vary in the details. One class of model assumes that the magnetar wind nebula is surrounded by a supernova ejecta \citep[e.g.][]{margalit18a,li2020}, while the other models assume that the magnetar wind nebula is expanding freely \citet{dai2017,yang2019}. In spite of the several attempts, details of the radio emission are still unclear. For example, the most recent model \citep{li2020}, while can accurately match the previously published flux density measurements ($\ge 1$ GHz), severely under-predicts the flux density at 400 and 600 MHz. 
While, in their current forms the models are not applicable, these models might be able to match the new data presented here. However this is outside the scope of this paper. The models which assume that the magnetar is not surrounded by the supernova ejecta can still probably explain the observations by lowering the self-absorption frequency, but the details are outside the scope of this work. However, it is evident that the low frequency observations presented here will play crucial role in pinpointing the actual origin of FRB 121102.

Recently, it has been suggested based on the variability timescales of the compact source associated with FRB121102, that its radius is $\sim0.01$ pc \citep{yang2020}. 
For getting a more robust handle on the radius, we model the spectrum and see what is the allowed range of magnetic field and radius so that all the available physical and observational constraints are satisfied.

We parametrised the electron injection spectrum using Eq. \ref{eq:elec_dis}. We have done a grid search in $\alpha$, radius and $\gamma_b$ and checked for which parameter values the observational constraints are satisfied. 
The value of $\alpha$ was varied between $0.01-10^{2.5}$. Radius of the source was varied between $0.001-0.65$ pc. $\gamma_b$ was varied between $10^3-10^6$. 
We assumed that the covering fraction is 1. Then for each ordered triplet, we calculated the total energy injected into the nebula, its age, $\epsilon_b$, $\epsilon_e$, $\nu_a$ and $\nu_m$ and then checked if they matched the constraints: a) $\epsilon_e+\epsilon_b \sim 1$ b) $E_{\rm{tot}}>6\times 10^{47}$erg \citep{wang2020}  c) Age is greater than 7 years. d) $\nu_a<0.4$ GHz, $\nu_m<1$ GHz

Based on this calculation, we conclude that the radius of the compact source is $\gtrsim 0.1$pc. In our calculations, the magnetic field of the nebula varied between $0.001-0.05$G. This is comparable to the characteristic field calculated in \citet{michilli2018}. We find that $\nu_a$ is always greater than $\nu_m$. In this light, it seems that the increasing flux density with increasing frequency between 0.4--1 GHz (shown by the black dashed line) is not the right model. However, more observations are needed to investigate whether these predictions are correct.

\subsection{Comparison between PTF10hgi and FRB121102 properties}

The qualitatively similar spectra of PTF10hgi and FRB121102 compact source indicates a common origin of the persistent radio emission.
However, the details of the two emission processes do not match. While the emission from FRB121102 can be modelled by a pulsar/magnetar wind nebula which is not surrounded by supernova ejecta, the same is not true in case of PTF10hgi. Such a scenario can only be produced after a supernova explosion if the neutron star gets a high kick and is ejected from the supernova. Assuming such a situation, we solve the relevant equations given in \citet{dai2017} and check for inconsistencies. We find that the obtained radius is $\sim 0.4$ pc, which is only possible if the wind velocity is of the order of 18000 km/s making this scenario highly improbable. Hence, we can safely suggest that this model is not applicable for PTF10hgi.  

In \S \ref{sec:PTF10hgi_properties} and \S  \ref{sec:FRB121102} we have shown that the radius of the continuum radio emitting region in PTF10hgi and FRB121102 is $\sim 0.04$pc and $\gtrsim 0.1$pc respectively. Assuming typical wind velocities in both the system, this result implies that the age of FRB121102 compact source is much larger than PTF10hgi. 
In \S \ref{sec:PTF10hgi_properties} we have shown that the average energy injection rate of PTF10hgi is $3 \times 10^{38}\,$erg$\,$s$^{-1}$. The average energy injection rate for FRB121102 is $\ge 2.7 \times 10^{39}\,$erg$\,$s$^{-1}$ \citep{wang2020}. 
{The energy injection rate depends on the exact details of the particle acceleration processes. At this point, the exact details of this process is not known, to the best of our knowledge. We hypothesise that the difference between the energy injection rate between PTF10hgi and FRB121102 stems from the efficiency with which the two magnetars can accelerate particles, which in turn can, in principle, vary between different sources.
However, since these energetic particles are the source of the FRBs, our analysis suggest that if there are radio bursts from PTF10hgi, they, on an average, will be significantly weaker energetically than that in FRB121102.} If detected, these bursts will bridge the gap between the galactic FRB from SGR 1935+2154 \citep{anderson2020} and FRB121102, whose properties agree very well with that of predictions from the magnetar scenario. Our measurements also suggest that the nebula of PTF10hgi is most probably being powered by the rotational energy of the magnetar residing inside the nebula which might also be powering the FRB121102 compact source \citep{wang2020}.


\section{Conclusion} \label{sec:conclusion}

Here, we present for the first time, the wideband spectrum of PTF10hgi ranging from 0.6-15 GHz. The flat spectrum between 1--15 GHz is very similar to that obtained in the pulsar wind nebulae in the Milky Way. Based on this and the fact { core dominated AGNs are very less likely to be present in a dwarf galaxy, we conclude that the radio emission is most probably originating from the magnetar wind nebula. Hence, in a way our results independently verify the suggestion that a magnetar was born during the explosion of PTF10hgi \citep{inserra13}. Our results suggest that the radio emission from PTF10hgi is most probably originating from the magnetar wind nebula and is very similar to the properties of the radio source associated with FRB121102. Although, this strengthens the connection between FRBs and Type I SLSN, a formal link between them will only be established if a FRB is detected from SLSN I like PTF10hgi. Hence searching for FRBs towards PTF10hgi will be very interesting. }



\section*{Data Availability}

Electronic image at all frequencies are provided in the Supplementary material.

\section*{Acknowledgements}
The authors acknowledge support of the Department of Atomic Energy, Government of India,under the project no. 12-R\&D-TFR-5.02-0700. 
PC acknowledges the support from the Department of Science and Technology via Swarna Jayanti Fellowship awards (DST/SJF/PSA-01/2014-15). We thank the staff of the Giant Metrewave Radio Telescope (GMRT) and the National Radio Radio Astronomy Observatory (NRAO) that made these observations possible. The Giant Metrewave  Radio  Telescope  is  run  by  the  National Centre for Radio Astrophysics of the Tata Institute of Fundamental Research. The National Radio Astronomy Observatory is a facility of the National Science Foundation operated under cooperative agreement by Associated Universities, Inc. SM and AB also thank Barun Maity (NCRA-TIFR) and Minhajur Rahaman (NCRA-TIFR) for insightfull discussions.





\bibliographystyle{mnras}
\bibliography{references} 

\newpage


\appendix

\section{Observations and Data analysis}

{
\subsection{uGMRT observations \& Data Analysis of PTF10hgi}

We observed PTF10hgi with the uGMRT band-4 (550--900 MHz) and band-5 (1000--1500 MHz) wide-band receivers to probe its low frequency radio emission. Band-5 observations were carried out on December 28, 2019, for 5 hours of total observing time ($\approx$ 4 hours on-source) with a 400 MHz band-width centred at 1200 MHz, under the proposal ID 37\_112. Radio sources 3C286 and J1557-000 were used as the flux calibrator and the phase calibrator, respectively. Observations in the uGMRT band-4 were carried out under the proposal DDTC114 using uGMRT \textit{director's discretionary time} on January 12 and February 14, 2020, with three hours of observing time in each day, which together provided $\approx$ 5 hours of on-source time. For the band-4 observations, radio sources 3C48 and 3C286 were used for flux and band-pass calibration while J1557-000 and J1649+123 were used for phase calibration.

For both the uGMRT frequency bands, basic data editing, gain and band-pass calibration were done using the classic AIPS package and subsequently, self-calibration was done using the CASA package \citep{casaref}. An automated RFI flagging software, aNKflag \citep{bera00}, was used for RFI excision at different stages of data analysis. Final imaging, in both the frequency bands, was done using the Briggs' weighting scheme \citep{briggs95} with \textit{robust} = 0, to reduce the deconvolution artefacts in the radio image. 

For uGMRT band-5, the full available band-width (1000--1400 MHz) was used to produce the radio continuum image centred at 1.2 GHz, which has a pixel size of $0.5$", a synthesized beam of 2.3"$\times$2.0" and an RMS noise of $\approx$ 9.1 $\mu$Jy. The target source PTF10hgi is clearly detected in this band at $\sim 7 \sigma$ significance with a radio continuum flux of 64 $\pm$ 9 $\mu$Jy (see figure \ref{fig:image1}). The band-4 image was made using 240 MHz of frequency bandwidth centred at 685 MHz. This image has a pixel size of $1^"$, a synthesised beam of $4.9^"\times 4.6^"$ and an RMS noise of $\approx$ 12 $\mu$Jy. No significant radio emission is detected from the target source in this image, as shown in figure \ref{fig:image1}.


\begin{figure*}
     \centering
     \includegraphics[trim={0 0.6cm 0.8cm 0},clip,scale=0.7]{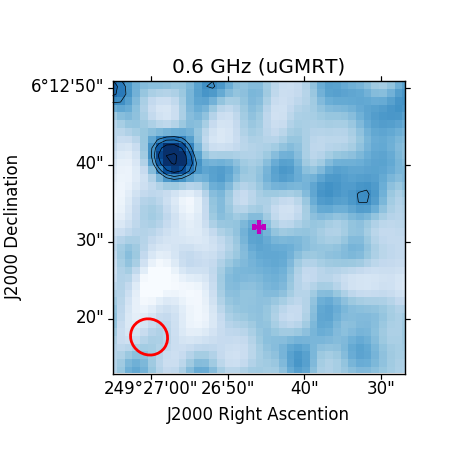}
     \includegraphics[trim={0.5cm 0.6cm 0.8cm 0},clip,scale=0.7]{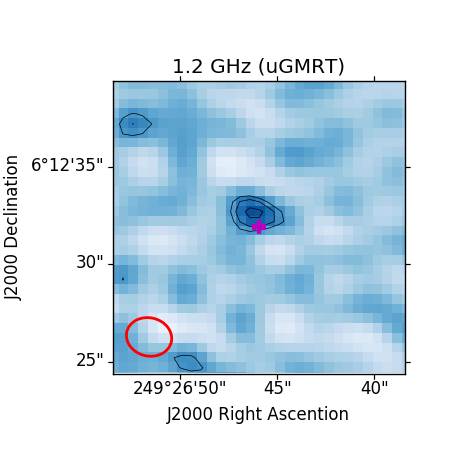}
     \includegraphics[trim={0cm 0.6cm 0.8cm 0},clip,scale=0.7]{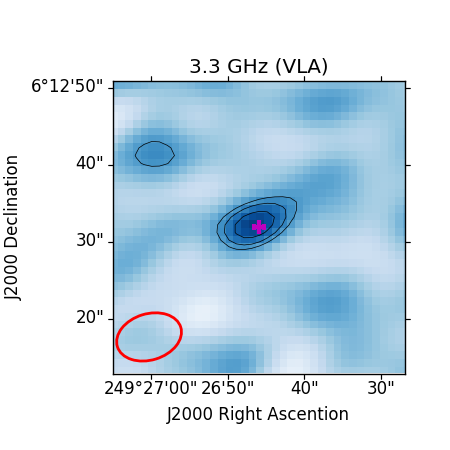}
     \includegraphics[trim={0.5cm 0.6cm 0.8cm 0},clip,scale=0.7]{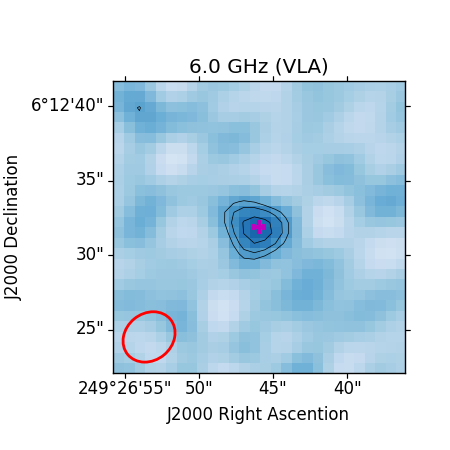}
     \includegraphics[trim={0 0.4cm 0.8cm 1.2cm},clip,scale=0.7]{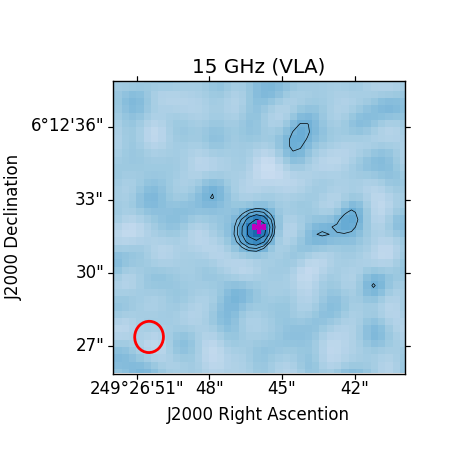}
     \caption{PTF10hgi images obtained at different bands.  The lowest contour level is at 2$\sqrt{2}\sigma$ and then increases in steps of $\times \sqrt{2}$. The red cross shown at the centre of the image shows the location of PTF10hgi (obtained from SLSN database \url{https://slsn.info/obj/PTF10hgi.html}). The red ellipse at the lower left of each panel shows the effective resolution.}
     \label{fig:image1}
 \end{figure*}

\subsection{VLA observations \& Data Analysis of PTF10hgi}

We observed PTF10hgi with the JVLA using the S (2--4 GHz), C (4--8 GHz) and Ku (12--15 GHz) band receivers.  For the S and C bands 8-bit samplers were used, whereas 3-bit samplers were used in case of Ku band. The observations were taken under the proposal codes 19B--347 and 20A--424 on 28th March, 28th February and 17th April, 2020 in the S, C and Ku bands respectively. 3C286 was used a flux calibrator for all observations. J1640--1220 was used as phase calibrator in the S and C band observations. J1658+0741 was used both as pointing calibrator and phase calibrator in the Ku band observation.
The data were calibrated using the VLA calibration pipeline. The calibrated data for the target was then split out and imaged using the \textsc{casa} task `tclean'. Final imaging, in both the frequency bands, was done using the Briggs' weighting scheme \citep{briggs95} with \textit{robust} = 0, to reduce the deconvolution artefacts in the radio image.  Details about the image and the flux density of the radio source associated with PTF10hgi is given in Table \ref{tab:source_properties}.

\begin{table*}
\caption{PTF10hgi image properties at different frequencies}
    \centering
    \begin{tabular}{|c|c|c|c|c|}
    \hline
    \hline
        Observing frequency (GHz) & Pixel scale & Restoring beam & Flux density ($\mu Jy$) & rms ($\mu Jy$) \\
    \hline    
         0.6 & $1^{''}$ & $4.9^{''}\times 4.6^{''}$ & $<36$ ($3\sigma$) & 12 \\
         1.2 & $0.5^{''}$ &  $2.3^{''}\times2.0^{''}$ & 64 & 9 \\
         3.3 & $1^{''}$ &  $8.6^{''}\times 6^{''}$ & 66 & 9 \\
         6 & $0.7^{''}$ & $3.7^{''}\times 3.1^{''}$ & 46 & 7 \\
         15 & $0.3^{''}$ & $1.3^{''}\times 1.2^{''}$ & 47 & 4 \\
    \hline     
    \end{tabular}
    \label{tab:source_properties}
\end{table*}

\subsection{uGMRT Archival Data Analysis of FRB121102}

The low frequency interferometric data for FRB121102 were downloaded from the uGMRT public data archive. The observations in band-3 (250--500 MHz) were carried out in July--August, 2017 (proposal ID: 32\_077, P.I.: S. Chatterjee) while the band-4 (550--900 MHz) observations were done in September, 2019 (proposal ID: DDTC090, P.I.: V. Gajjar). After initial gain and band-pass calibration, self-calibration and interferometric imaging were done following the standard procedure, very similar to what has been done for analysing the uGMRT data for PTF10hgi. The pixel scale for band 3 and band 4 images are $1.3^{''}$ and $1^{''}$ respectively. The restoring beams are $6.1^{''}\times 5.1^{''}$ and $5.2^{''}\times4.5^{''}$ respectively.
The compact persistent radio source associated with FRB121102 was detected in both the uGMRT frequency bands. Its observed flux densities are 215$\pm$37 $\mu$Jy at 433 MHz and 231$\pm$22 $\mu$Jy at 668 MHz. The uGMRT images are shown in Fig. \ref{fig:FRB121102_images}.}

 \begin{figure*}
     \centering
     \includegraphics[trim={0 0.6cm 0.8cm 0},clip,scale=0.9]{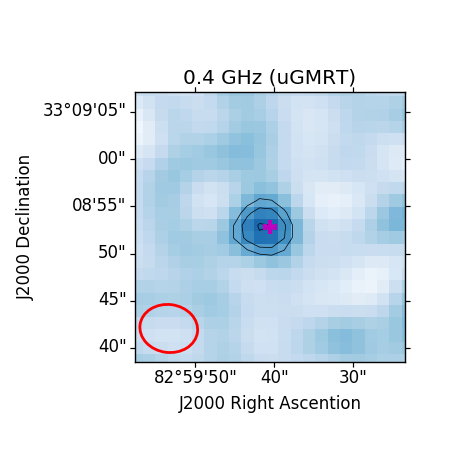}
     \includegraphics[trim={3.2cm 0.6cm 0.8cm 0},clip,scale=0.9]{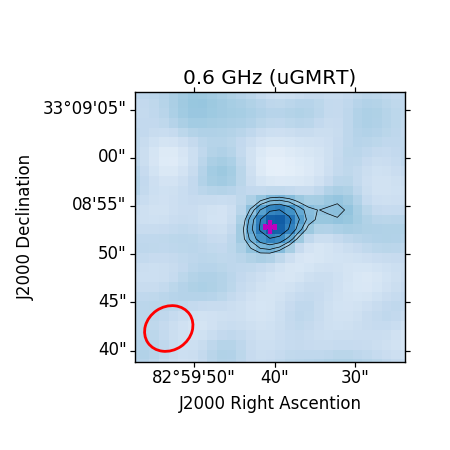}
     \caption{Shows the images of the persistent radio source associated with FRB121102 at uGMRT band 3 and band 4. Left panel: Band 3 image Right panel: Band 4 image. The lowest contour level is at 2$\sqrt{2}\sigma$ and then increases in steps of $\times \sqrt{2}$.  The red ellipse at the lower left of each panel shows the effective resolution. The magenta cross shows the location of the persistent source reported in \citet{chatterjee2017}.}
     \label{fig:FRB121102_images}
 \end{figure*}

\label{lastpage}

\end{document}